\documentclass[10pt,a4paper]{article}
\pdfoutput=1
\usepackage[english]{babel}
\usepackage{longtable}
\usepackage{amsmath,amsfonts,amssymb,bm}\interdisplaylinepenalty=2500

\usepackage{graphicx,color,rotating,epstopdf}
\graphicspath{{figs/}}					
\def\PrintGraphicFileName{1}			
\newcommand{\namedgraphics}[3]{
   \parbox{#3}{\ifnum\PrintGraphicFileName>0\rotatebox{90}{\smash{\ttfamily\scriptsize\raisebox{0.8em}{#2}}}\fi%
   \hspace*{\fill}\includegraphics[scale=#1]{#2}\hspace*{\fill}}}

\title{Phase and frequency noise metrology}

\author{E. Rubiola,  V. Giordano, K. Volyanskiy, L. Larger\\
\small web page \texttt{http://rubiola.org}
\\[4em]\includegraphics[width=0.35\textwidth]{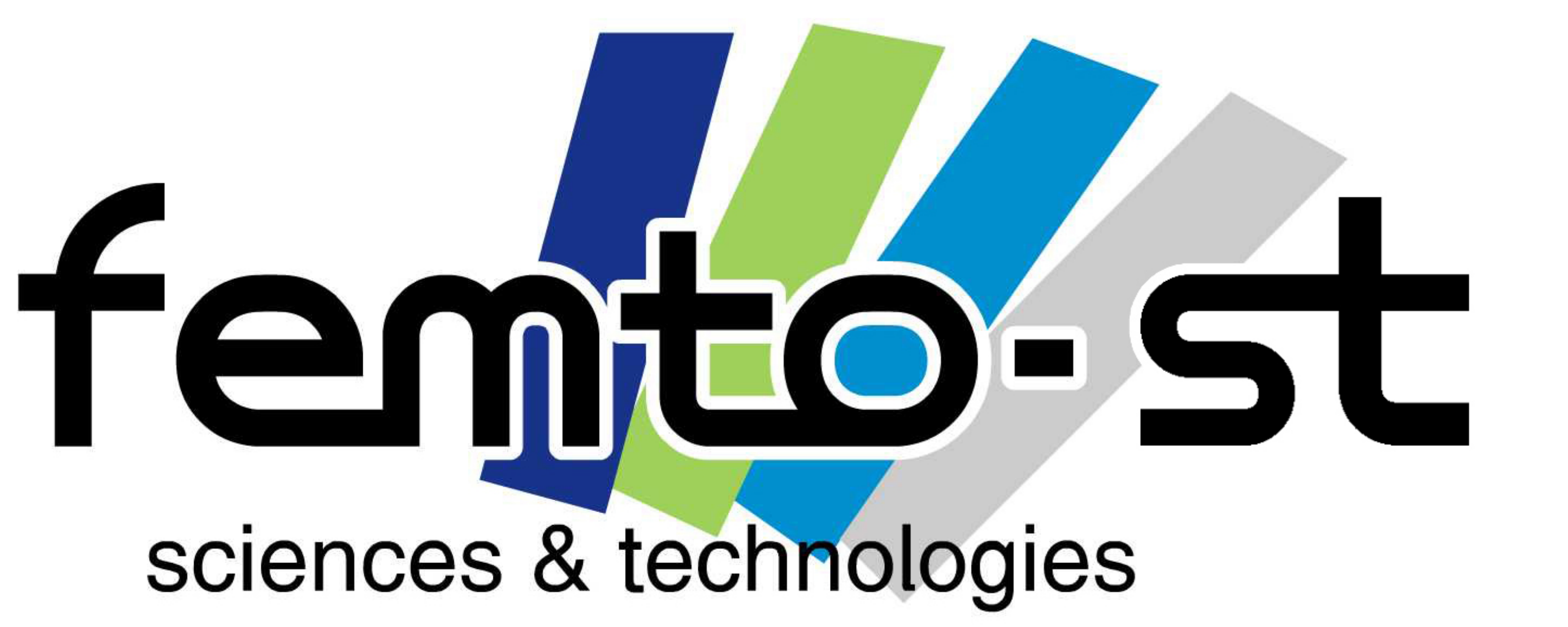}\\[0.5em]
\small FEMTO-ST Institute\\[-0.5ex]
\small CNRS and Universit\'e de Franche Comt\'e, 
\small Besan\c{c}on, France\\[1.5em]}
\date{\small\today}
\def\myheaders{E. Rubiola \& al., Phase and frequency noise metrology.\hfill Nov.\,30, 2008\quad}
\begin{document}
\maketitle

\begin{abstract}
As a consequence of a general trend in the physics of oscillators and clocks towards optics, phase and frequency metrology is rapidly moving to optics too.  Yet, optics is not replacing the traditional radio-frequency (RF) and microwave domains.  Instead, it adds tough challenges.

Precision frequency-stability measurements are chiefly based on the measurement of phase noise, which is the main focus of this article.
Major progress has been achieved in two main areas.  The first is the extreme low-noise measurements, based on the bridge (interferometric) method \cite{ivanov98uffc,rubiola1999rsi} in real time or with sophisticated correlation and averaging techniques \cite{rubiola2000rsi-correlation,rubiola2002rsi-matrix}.  The second is the emerging field of microwave photonics, which combines optics and RF/microwaves.  This includes the femtosecond laser, the two-way fiber links \cite{foreman2007rsi-optical-link}, the noise measurement systems based on the fiber \cite{volyanskiy2008josab-optical-fiber} and the photonic oscillator \cite{yao96josab,yao00jlt}.  Besides, the phenomenology of flicker ($1/f$) noise is better understood, though the ultimate reasons are still elusive.
\end{abstract}

\clearpage
\tableofcontents

\section{A quick look to high-sensitivity measuements}\label{sec:fsm-intro}
\begin{figure}[b]
\centering\namedgraphics{0.40}{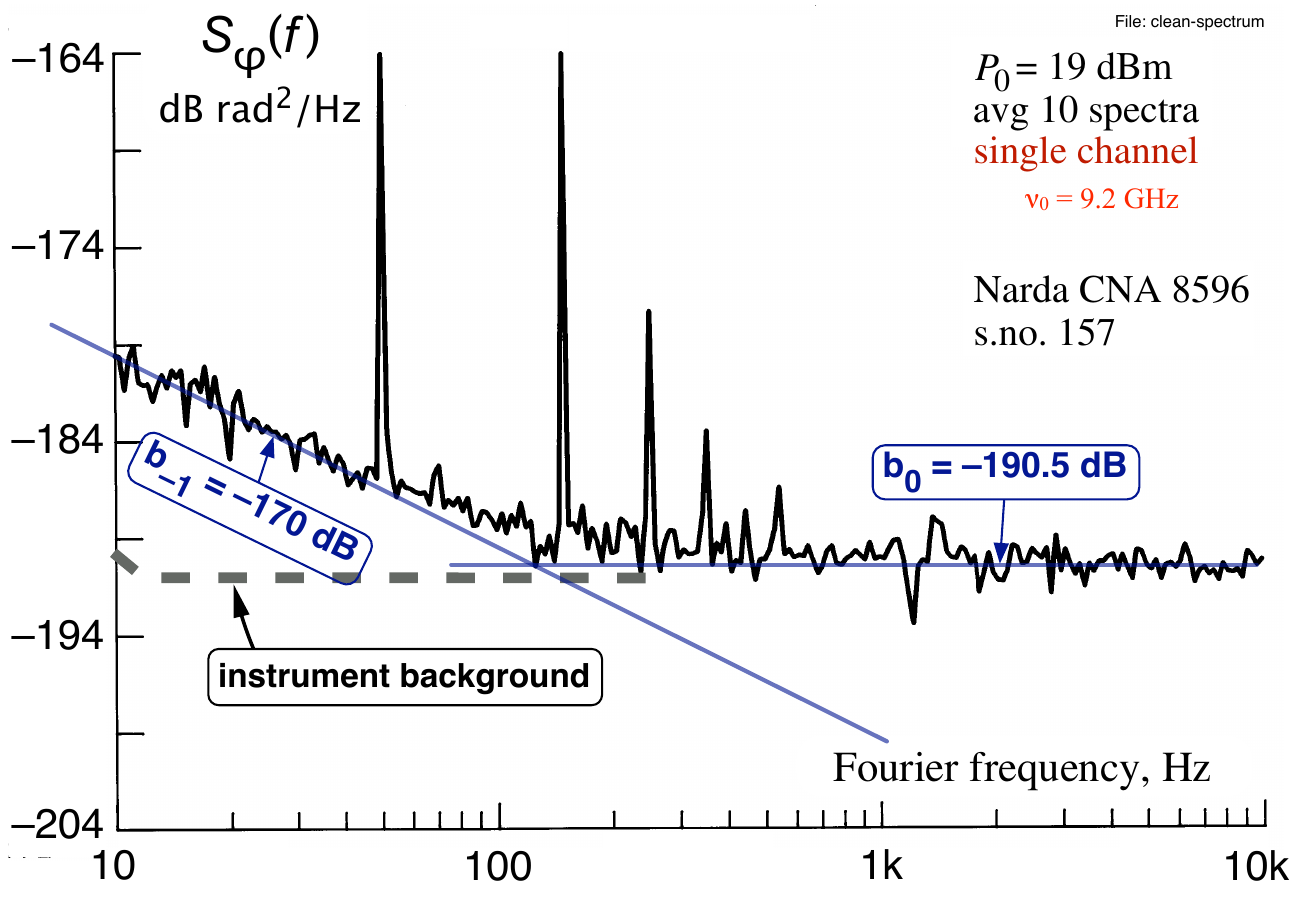}{\columnwidth}
\caption{Phase noise spectrum of a ferrite circulator measured at 9.2 GHz \cite{rubiola2004uffc-circulators}.}
\label{fig:fsm-clean-spectrum}
\end{figure}
Before getting through the basic principles, let us look at the phase noise spectrum of Figure~\ref{fig:fsm-clean-spectrum}, approximated with the power-law $S_\varphi=\sum_{i}b_i/f^i$.  The device under test is a microwave circulator with the output taken at the isolation port.  Isolation results from the destructive interference between two counter-propagating modes in the ferrite bulk.  This configuration is the same used in the well known Pound oscillator \cite{pound46rsi}.
The experimentalist familiar with phase noise should notice the following unusual facts.
\begin{enumerate}
\item The low $1/f$ noise of the DUT, $10^{-17}/f$ $\mathrm{rad^2/Hz}$, measured without need of correlation.
\item The low background noise, $10^{-18}/f+9{\times}10^{-20}$ $\mathrm{rad^2/Hz}$.
\item The low level of the residual of the mains, 50 Hz and odd multiples.  The even-order harmonics are so small that they are not visible.
\item The smoothness of the plot, achieved with a small number of averaged samples ($m=10$).
\end{enumerate}
The background noise deserves more attention.  The conversion between spectrum and Allan variance, which is a simple mathematical process, can be done with to any physical quantity.  The formula $\sigma^2(\tau)=2\ln(2)\,h_{-1}$ applies to flicker $S(f)=h_{-1}/f$.  
\begin{figure}[t]
\centering\namedgraphics{0.4}{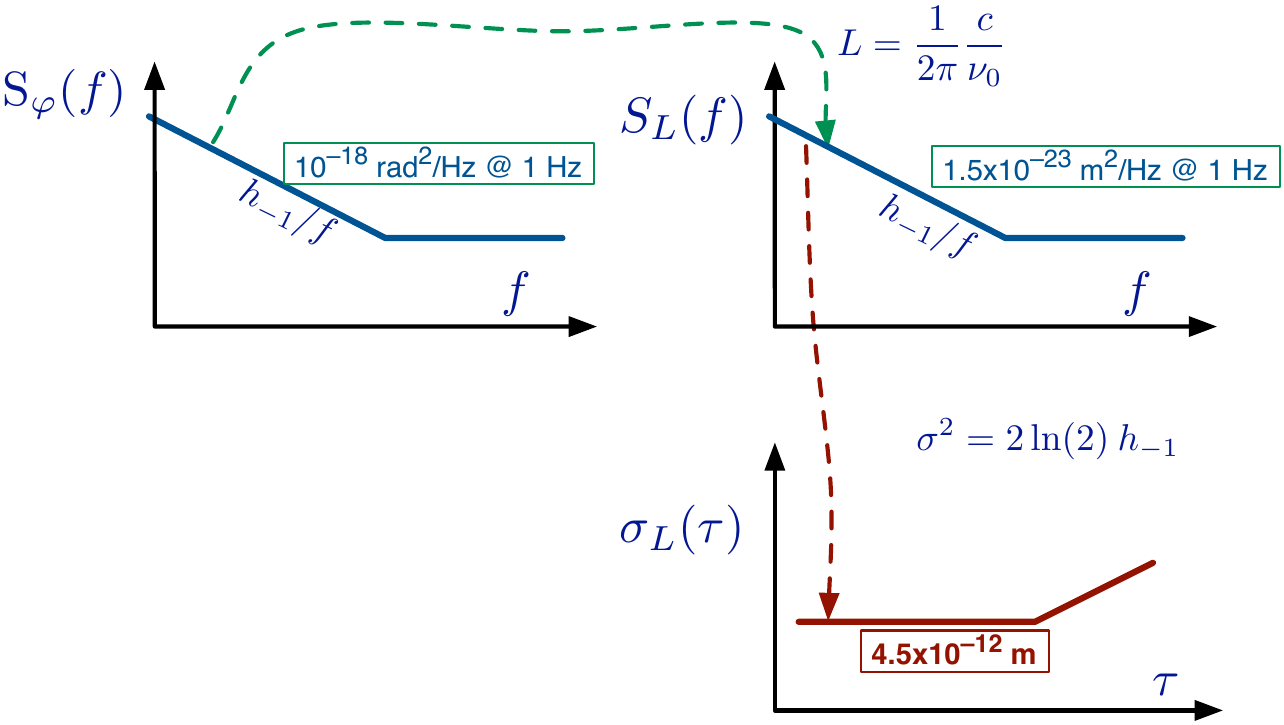}{\columnwidth}
\caption{Phase noise converted into mechanical stability.}
\label{fig:fsm-mechanical-stability}
\end{figure}
Therefore, after converting the phase spectrum $S_\varphi=b_{-1}/f$ into a length-fluctuation spectrum $S_L=h_{-1}/f$, the Allan deviation is $\sigma_L(\tau)=4.5{\times}10^{-12}$ m (Fig.~\ref{fig:fsm-mechanical-stability}).
This high stability requires to re-think the mechanical design of electronics.  That said, a resolution of parts in $10^{-14}$ m is common in the field of tunnel and atomic-force microscopy \cite{sarid:microscopy}.  Also the femtosecond laser owes its success to the unexpected mechanical stability of the optical assembly.

\section{Bridge measurements}\label{sec:fsm-bridge}
\begin{figure}[t]
\centering\namedgraphics{0.5}{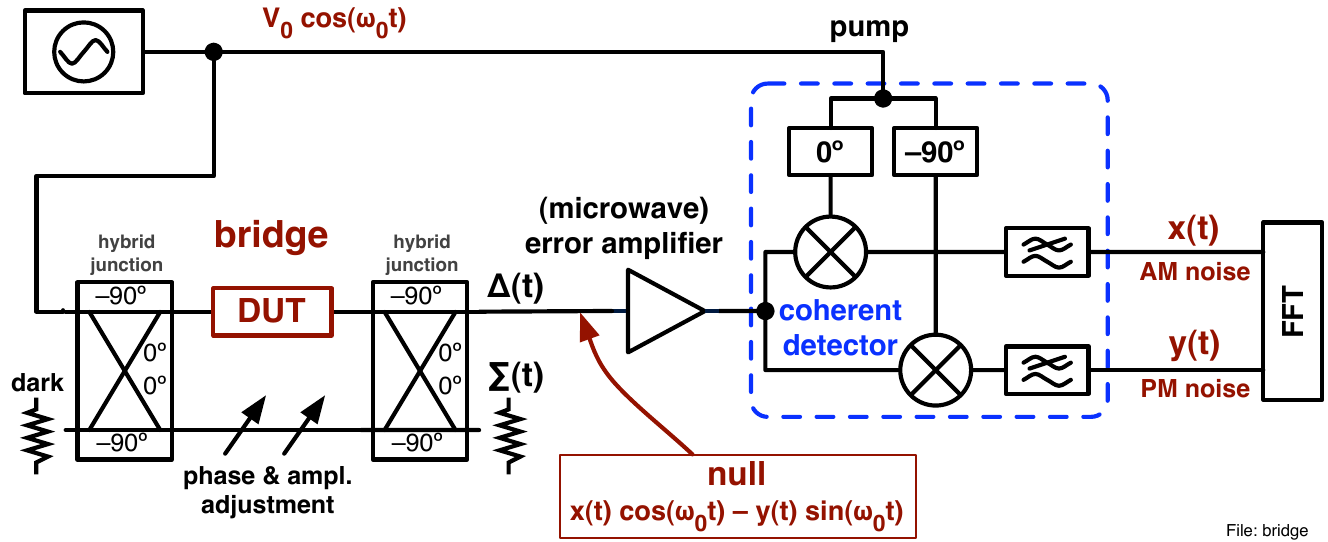}{\columnwidth}
\caption{Bridge (interferometric) method for the measurement of phase noise.}
\label{fig:fsm-bridge}
\end{figure}

The instrument, shown in Fig.~\ref{fig:fsm-bridge}, is based on the carrier suppression by sum of an equal and opposite signal. The null contains only the noise sidebands of the DUT, which are amplified and down-converted to dc by coherent detection \cite{viterbi:communication}.  The in-phase signal $x(t)$ is proportional to the normalized-amplitude noise (AM noise) $\alpha(t)$, the quadrature signal $y(t)$ to the PM noise $\varphi(t)$
The following ideas makes this scheme a fortunate choice. 
\begin{enumerate}
\item The $1/f$ noise of the passive components the bridge is made of is dramatically reduced by separating coarse and fine the adjustment of the null.  The coarse adjustment relies on by-step and fixed-value components.  The fine adjustment, though more noisy because of the contact fluctuations, has low weight.

\item Close-in flicker results from up-conversion of the near-dc $1/f$ noise pumped by the carrier (cf.\ Sec.~\ref{sec:fsm-devices}).  Here the error amplifier sees only the fluctuation of the null.  Power is too low for the up-conversion process.

\item The noise floor is improved significantly by microwave amplification of the noise sidebands before the mixer loss.  

\item Microwave amplification of the noise sidebands before detecting reduces the residuals of the mains because the dc electronics is highly sensitive to these fields, the microwave section is not.

\item The in-phase/quadrature detection is a fully-linear Cartesian process.  The dc component of $x(t)$ and $y(t)$ can be used to control the null in closed loop.  This is recommended when the experiment lasts more than some half an hour.
\end{enumerate}
The bridge has been used successfully to measure the flicker of ferrites \cite{rubiola2002rsi-matrix,rubiola2004uffc-circulators,woode1998mst-isolators} ($b_{-1}\approx10^{-17}$ $\mathrm{rad^2/Hz}$, either HF/VHF and microwave carrier) and photodetectors \cite{rubiola2006mtt-photodiodes} ($b_{-1}\approx10^{-12}$ $\mathrm{rad^2/Hz}$, 10 GHz modulation).  These measurements are out of reach for other methods, the first because of the extremely low noise, the second because of the low microwave power (order of 10 $\mu$W).

The photodetector $1/f$ noise is relevant in OEOs (Sect.~\ref{sec:fsm-oeo-noise}) and in optical-fiber links \cite{foreman2007rsi-optical-link}.  In the link, the detector noise can not be removed with the two-way method.  The value $b_{-1}=10^{-12}$ $\mathrm{rad^2/Hz}$ at 10 GHz sets the stability limit to $\sigma_x\simeq1.9{\times}10^{-17}$ s (Allan deviation). 
In the same condition the fractional-frequency stability of frequency transfer is $\sigma_y\simeq4.2{\times}10^{-17}/\tau$, not far from the performance of short-range links.  This is seen using $S_y=(f^2/\nu_0^2)\,S_\varphi$ and converting $S_y$ into Allan deviation.

\section{Correlation measurements}\label{sec:fsm-correl}
\begin{figure}[t]
\centering\namedgraphics{0.5}{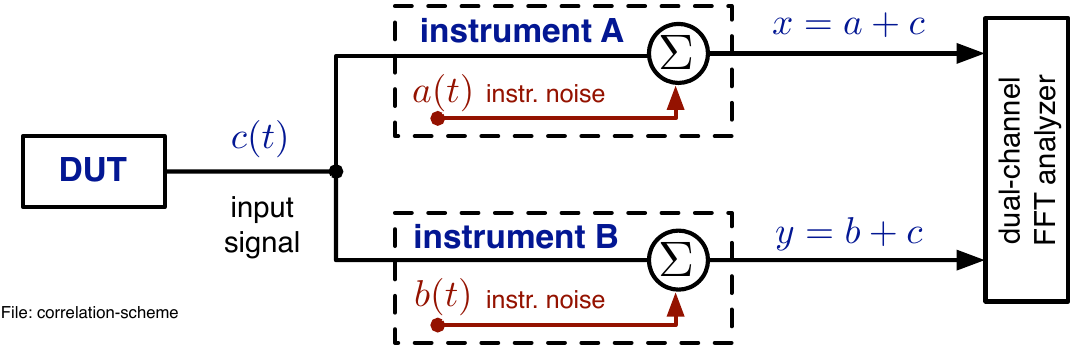}{\columnwidth}
\caption{Basic correlation measurement.}
\label{fig:fsm-correlation-scheme}
\end{figure}

Two equal instruments measure the same DUT, as shown in Fig.~\ref{fig:fsm-correlation-scheme}.  This figure also introduce the main symbols.  Notice that $x(t)$ and $y(t)$ are \emph{not} the same thing as in Sec.~\ref{sec:fsm-bridge}.
We denote the Fourier transform with the uppercase of the time-domain function, the complex conjugate with $\ast$, and the average of $m$ samples with $\left<\:\right>_m$.  Thanks to the Wiener-Kinchine theorem, the cross-spectrum is $S_{yx}=YX^\ast$.  Hence
\begin{align}
S_{yx}
&= \left< [B+C] \times [A+C]^\ast \right>_m\\
&= CC^\ast + O(1/m)~.
\end{align}
Thus, $S_{yx}$ converges to the spectrum $S_c=CC^\ast$ of the DUT, with a residual random term of the order of $1/m$\@.

In the laboratory practice, it is often convenient to display $|\left<\Re\{S_{yx}\}\right>_m|$ because it is the minimally-biased always-positive estimator, assuming that $S_c$ is real.
With white noise we use the ergodicity principle to access the ensemble (a part of) by sweeping the frequency.  Thus we have access to the average and to the standard deviation of $|\left<\Re\{S_{yx}\}\right>_m|$.  
With small $m$, the single-channel noise is still not rejected.  This means that the term $O(1/m)$ is larger than the DUT noise.  In this condition, the deviation of $|\left<\Re\{S_{yx}\}\right>_m|$ is almost equal to the average.  In logarithmic scale, the spectrum looks as a band of constant thickness that moves downwards on the analyzer display.  Increasing $m$, the term $O(1/m)$ becomes lower than the DUT noise.  Thus, the deviation becomes  smaller than the average.  The track attains a constant value and shrinks.
\begin{figure}[t]
\centering\namedgraphics{0.64}{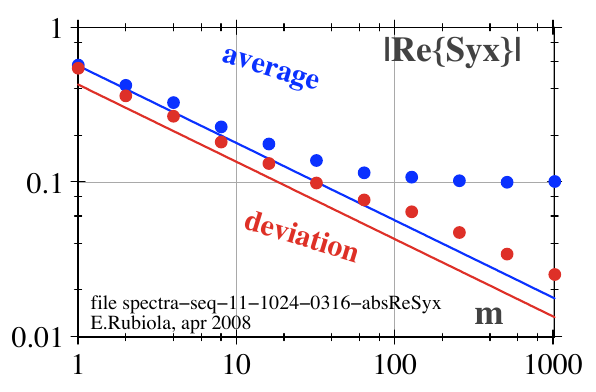}{\columnwidth}
\caption{Convergence of $|\left<\Re\{S_{yx}\}\right>_m|$ to $S_c$ in the case of the white noise floor.  In the example, the DUT noise is 10 dB lower than the single-channel background of the instrument.}
\label{fig:fsm-correlation-m-law}
\end{figure}
This is shown in Fig.~\ref{fig:fsm-correlation-m-law}, where for $m>32$ the deviation of $|\left<\Re\{S_{yx}\}\right>_m|$ becomes significantly lower than the average.

Interestingly, the fact that the deviation/average ratio that decreases for $m$ larger than a threshold validates the measurement mathematically, without need of checking on the instrument background.

\begin{figure}[t]
\centering\namedgraphics{0.70}{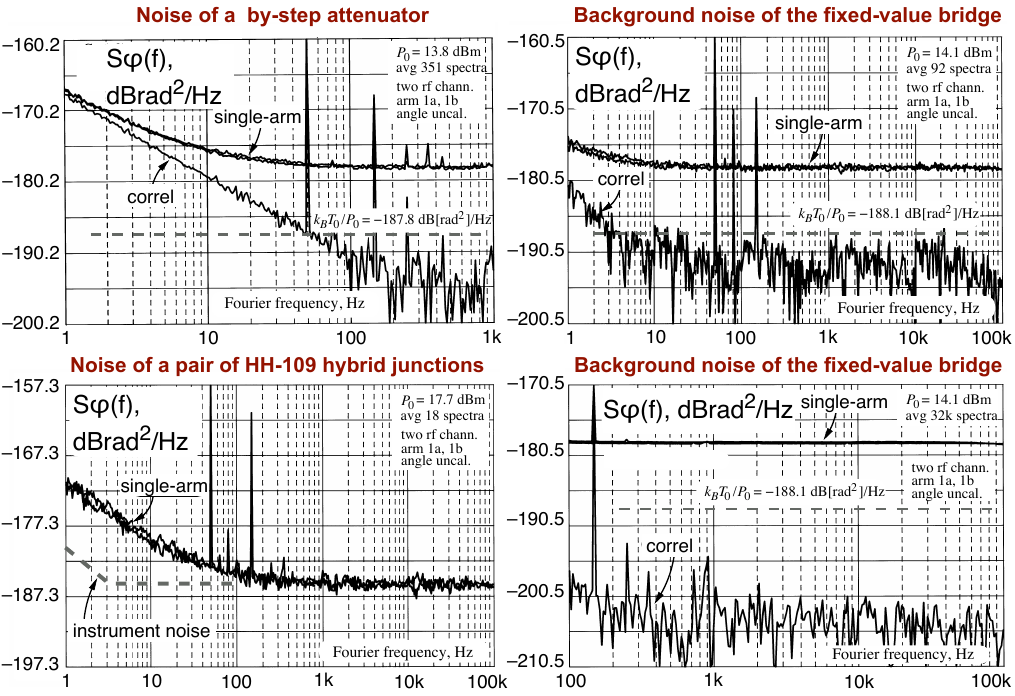}{\columnwidth}
\caption{Measurement of low-phase-noise devices.  The background noise of the instrument is measured in the same conditions, but with different frequency span and number of averaged spectra.}
\label{fig:fsm-sphi-spectra}
\end{figure}
Correlation is an old idea \cite{hanbury-brown52nat,vessot64nasa,walls76fcs}, which could be used routinely in phase noise measurements only with the FFT analyzer \cite{wwalls92fcs}.
The use of correlation on a bridge instrument enables the measurement of some devices that could not be measured before \cite{rubiola2002rsi-matrix}. 
Figure~\ref{fig:fsm-sphi-spectra} shows some examples.

\section{Measurement of AM noise and RIN spectra}\label{sec:fsm-rin}
\begin{figure}[t]
\centering\namedgraphics{0.63}{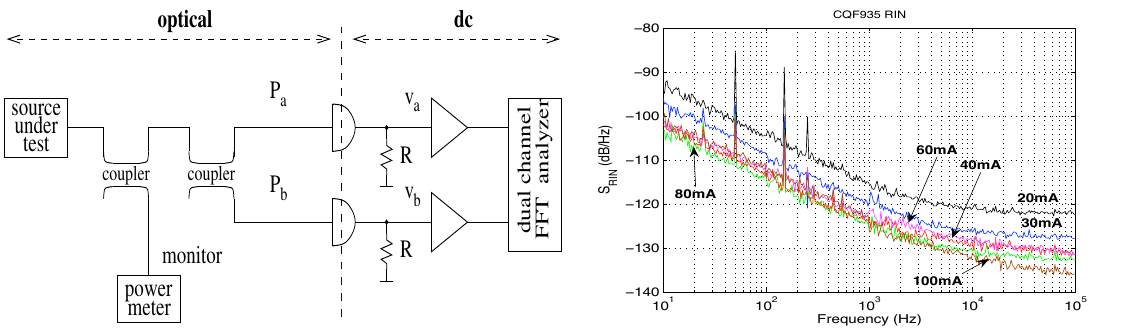}{\columnwidth}
\caption{Measurement of the laser RIN spectrum.}
\label{fig:fsm-rin}
\end{figure}
In the case of PM noise, the single-channel background is usually measured by removing the DUT and by replacing it with a short cable, which is virtually noise free.  In the case of AM noise and RIN, the DUT can \emph{not} be removed because the DUT is the only source of signal.  Thus, the validation of the instrument requires a reference of suitably low amplitude noise, which is not available in general.  Correlation solves the problem because the convergence of the spectrum to the DUT noise can be assessed mathematically through the deviation/average ratio, as explained in Sec.~\ref{sec:fsm-correl}.
Figure~\ref{fig:fsm-rin} shows an example of laser RIN measurement.

\section{Flicker in electronic and optical devices}\label{sec:fsm-devices}
Though the ultimate reasons for the $1/f$ noise are still elusive, the mechanism of the close-to-the-carrier flicker in RF/microwave devices is simple.
Understanding it starts from the simple observation that 
\begin{itemize}
\item near-dc flicker exists per se, though often made in accessible by an output filter.
\item the output spectrum is white in the microwave band,
\item close-in noise appears only in the presence of a carrier.
\end{itemize} 
\begin{figure}[t]
\centering\namedgraphics{0.65}{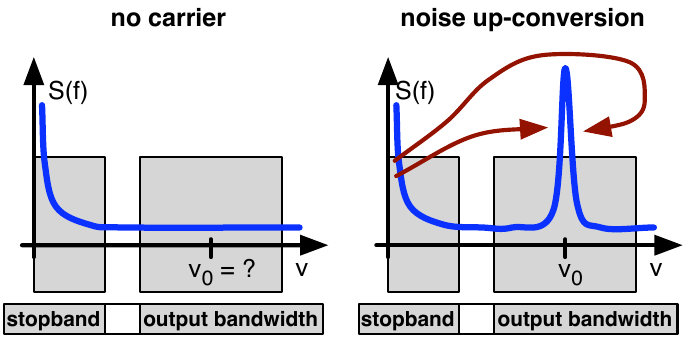}{\columnwidth}
\caption{Flicker up-conversion in electronic devices.}
\label{fig:fsm-noise-upconversion}
\end{figure}
The reason is that close-in noise is brought from dc to the vicinity of the carrier by parametric modulation, which at first approximation is linear (Fig.~\ref{fig:fsm-noise-upconversion}).  
This simple model accounts for the following experimental facts about the $1/f$ noise spectra\cite[Chap.\,2]{rubiola2008cambridge-leeson-effect}.
\begin{itemize}
\item The noise is about independent of the carrier power.
\item The noise of cascaded devices is the sum of the individual spectra, regardless of the order in which the devices are chained. 
\item When equal devices are connected in parallel, the noise is that of one device divided by the number of devices. 
\end{itemize}
The typical value of the flicker parameter $b_{-1}$ in amplifiers is of $10^{-14}$ $\mathrm{rad^2/Hz}$ for bipolar RF units, of $10^{-12}$ $\mathrm{rad^2/Hz}$ for SiGe microwave units, and of $5{\times}10^{-11}$ $\mathrm{rad^2/Hz}$ for HBT microwave units; and of $10^{-12}$ $\mathrm{rad^2/Hz}$ for high-speed photodetectors.

\section{Noise in OEOs}\label{sec:fsm-oeo-noise}
The opto-electronic oscillator (OEO) is an appealing alternative for high spectral purity microwave oscillators \cite{yao96josab,yao00jlt}. 
\begin{figure}[t]
\centering\namedgraphics{0.6}{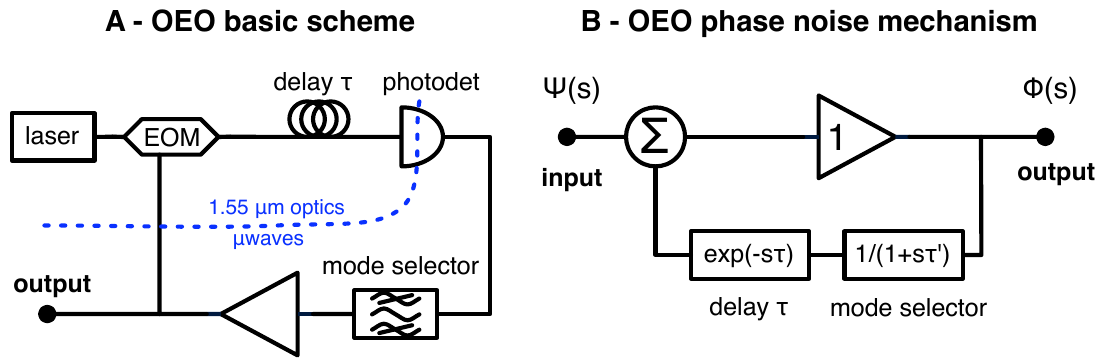}{\textwidth}
\caption{Basic scheme and phase-noise model in the OEO.}
\label{fig:fsm-oeo-scheme}
\end{figure}
Figure~\ref{fig:fsm-oeo-scheme} shows the basic scheme of an optical-fiber OEO and its phase noise model\cite[Chap.\,5]{rubiola2008cambridge-leeson-effect}.  Microwave oscillation can take place at any frequency multiple of $1/\tau$. A bandpass filter is necessary to select a single frequency.
In Fig.~\ref{fig:fsm-oeo-scheme}\,B all the signals are the Laplace transform of the \emph{phase fluctuation} of the microwave signal or modulation. 
Non-linearity necessary in real oscillator for the power not to decay or diverge.
Using amplitude and phase as the base representation, all the non-linearity goes in the amplitude, the phase is linear.  
The underlying physical fact is that time can not be stretched\footnote{This may not be true in the presence of strong non-linearity.}.  As a relevant consequence, phase noise is additive.  This eliminates the difficulty inherent in the parametric nature of some noise processes, like flicker.  
The input $\Psi(s)$ is the phase noise of all the oscillator components. The output $\Phi(s)$ is the phase noise at the oscillator output.  The amplifier gain is exactly equal to one because the amplifier repeats the phase of the input.
Elementary feedback theory yields to the phase-noise transfer function $\mathrm{H}(s)=1/[1-\mathrm{B}(s)]$, where $\mathrm{B}(s)$ represents the delay and the low-pass equivalent of the mode selector.  The function $|\mathrm{H}(jf)|^2$ is an extension of the Leeson model \cite{leeson66pieee}.

\begin{figure}[t]
\centering\namedgraphics{0.71}{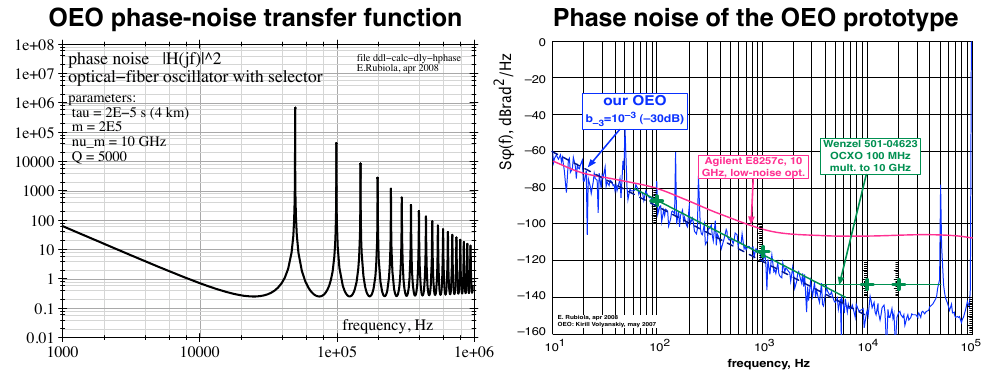}{\textwidth}
\caption{Examples of OEO phase-noise transfer function and spectrum.}
\label{fig:fsm-oeo-spectra}
\end{figure}
Figure \ref{fig:fsm-oeo-spectra} shows an example of transfer function $|\mathrm{H}(jf)|^2$ and of phase noise spectrum with 4 km optical fiber, i.e., with a delay of 20 $\mu$s.  The peaks of $|\mathrm{H}(jf)|^2$ at $n/\tau$ (50 kHz, 100 kHz, etc.) are due to the contiguous microwaves modes kept below the oscillation threshold by the mode selector.  The spectrum is measured with two delay lines used as the frequency discriminator, taking the average cross-spectrum to enhance the sensitivity \cite{volyanskiy2008josab-optical-fiber}. The $1/f^3$ noise is no more than 2 dB higher than the theoretical value, evaluated by putting the $1/f$ noise of the electronics in the Leeson formula.  The phase noise $10^{-3}/f^3$ (frequency flicker) is equivalent to the Allan deviation $\sigma_y=3.7{\times}10^{-12}$.

\def\bibfile#1{../../bibliography/#1}
\addcontentsline{toc}{section}{References}
\bibliographystyle{amsalpha}
\bibliography{\bibfile{ref-short},\bibfile{references},\bibfile{rubiola}}

\end{document}